\documentclass[10pt,a4paper]{article}
\usepackage{amsmath}
\usepackage{amssymb}
\usepackage{epsfig}

\newtheorem{thm}{Theorem}

\def\sn{\mathop{\rm sn}\nolimits}
\def\cn{\mathop{\rm cn}\nolimits}
\def\dn{\mathop{\rm dn}\nolimits}

\title{
On a Functional Equation of Ruijsenaars
}

\author{J.G.B. Byatt-Smith\thanks{E-mail:Byatt@ed.ac.uk}\ \ and
H. W. Braden\thanks{E-mail:hwb@ed.ac.uk}\\
\normalsize
\em Department of Mathematics and Statistics,\\
\normalsize
\em The University of Edinburgh, \\
\normalsize
\em Edinburgh, UK \\
}

\date{Submitted December, 2001}

\begin{document}

\renewcommand{\thepage}{}
\begin{titlepage}

\maketitle
\vskip-9.5cm
\hskip10.4cm
\vskip8.8cm

\begin{abstract}
We obtain the general solution of the functional equation
\begin{equation*}
\sum\limits_{\substack{ I\subseteq \{1,2,\ldots, n\} \\ |I|=k}}
\Bigg( \prod_{\substack{ i\in I \\ {j\not\in I}}} h(x_j-x_i)
h(x_i-x_j-i\beta)-
\prod_{\substack{ i\in I \\ {j\not\in I} }} h(x_i-x_j)
h(x_j-x_i-i\beta)\Bigg)=0.
\end{equation*}
This equation, introduced by Ruijsenaars, guarantees the commutativity of
$n$ operators associated with the quantum Ruijsenaars-Schneider models.
\end{abstract}

\begin{flushleft}
\end{flushleft}

\begin{flushleft}
\end{flushleft}

\vfill
\end{titlepage}
\renewcommand{\thepage}{\arabic{page}}

\section{Introduction}
The purpose of this paper is to investigate the functional equation
\begin{equation}
\sum\limits_{\substack{ I\subseteq \{1,2,\ldots, n\} \\ |I|=k}}
\Bigg( \prod_{\substack{ i\in I \\ {j\not\in I}}} h(x_j-x_i)
h(x_i-x_j-i\beta)-
\prod_{\substack{ i\in I \\ {j\not\in I} }} h(x_i-x_j)
h(x_j-x_i-i\beta)\Bigg)=0.
\label{qgencons}
\end{equation}
Here $\beta$ is an arbitrary positive number and the sum is over all 
subsets with $k$ elements.
As we will describe shortly,
this equation underlies the quantum integrability of the Ruijsenaars-Schneider
models. We will establish
\begin{thm}The general solution of the functional equation (\ref{qgencons})
analytic in a neighbourhood of the real axis with either a simple pole at the 
origin or an array of such poles at $np$ on the real axis ($n\in\mathbb{Z}$) 
is given by
\begin{equation}
h(x)=b\,\dfrac{\sigma(x+\nu)}{\sigma(x)\, \sigma(\nu)} \, e\sp{\alpha x}.
\label{eqsol}
\end{equation}
\end{thm}

Before turning to the proof of this theorem let us place this work in context.
Some years ago Ruijsenaars and Schneider \cite{RS} initiated the
investigation of mechanical models obeying the Poincar\'e algebra
\begin{equation}
\{H,B\}=P ,\qquad \{P,B\}=H,\qquad \{H,P\}=0.
\label{poincare}
\end{equation}
Here $H$ will be the Hamiltonian of the system generating time-translations,
$P$ is a space-translation generator and $B$ the generator of boosts.
Their study was motivated in part by seeking mechanical models that described
soliton interactions.
The models they discovered were found to posses other nice features: they were
in fact integrable and a quantum version of them naturally existed.
Ruijsenaars and Schneider began with the ansatz
$$
H=\sum_{j=1}\sp{n}\cosh p_j\, \prod_{k\ne j}f(x_j-x_k)
,\qquad
P=\sum_{j=1}\sp{n}\sinh p_j\, \prod_{k\ne j}f(x_j-x_k),\qquad
B=\sum_{j=1}\sp{n}x_j .  $$
With this ansatz and the canonical Poisson bracket $\{p_i,x_j\}=\delta_{ij}$
the first two Poisson brackets of (\ref{poincare}) involving the boost
operator $B$ are automatically satisfied.
The remaining Poisson bracket is then
\begin{multline*}
\{H,P\}=-\sum_{j=1}\sp{n}\partial_j \prod_{k\ne j}f\sp2(x_j-x_k) \\
-\frac{1}{2}\sum_{j\ne k}\cosh(p_j-p_k)\, \prod_{l\ne j}f(x_j-x_l)
\prod_{m\ne k}f(x_k-x_m)\Big(\partial_j\ln f(x_k-x_j)+\partial_k\ln f(x_j-x_k)
\Big)
\end{multline*}
and for the independent terms proportional to $\cosh(p_j-p_k)$ to vanish
we require that $f'(x)/f(x)$ be odd.
This entails that $f(x)$ is either even or
odd\footnote{Ruijsenaars and Schneider assume $f(x)=f(-x)$.}
and in either case $F(x)=f\sp2(x)$ is even.
Supposing that $f(x)$ is so constrained, then the final Poisson bracket is
equivalent to the functional equation
\begin{equation}
\{H,P\}=0 \Longleftrightarrow
\sum_{j=1}\sp{n}\partial_j \prod_{k\ne j}f\sp2(x_j-x_k)=0.
\label{functional}
\end{equation}
Observe that upon dividing (\ref{qgencons}) by $\beta$ and letting 
$\beta\rightarrow0$ this yields  (\ref{functional}) with $F(x)=h(x)h(-x)$
when $k=1$.

For $n=3$ equation (\ref{functional}) may be written in the form
\begin{equation}
\begin{vmatrix}
1              & 1              & 1              \\
F(x)           & F(y)           & F(z)           \\
F\sp{\prime}(x)& F\sp{\prime}(y)& F\sp{\prime}(z)\\
\end{vmatrix}
=0,\quad\quad  x+y+z=0,
\label{detdiff}
\end{equation}
where $F(x)=f\sp2(x)$. Ruijsenaars and Schneider \cite{RS}
showed that $F(x)=\wp(x)+c$
satisfies (\ref{detdiff}) and further satisfies (\ref{functional}) for all $n$.
This same functional equation (without assumptions on the parity of the
function $F(x)$) has arisen in several settings related to integrable systems.
It arises when characterising quantum mechanical potentials whose ground
state wavefunction (of a given form) is factorisable \cite{cal, suth, Gu}
.\footnote{As an aside we remark that the delta function
potential $a\delta \left( x\right) $
of many-body quantum mechanics on the line, which has a factorisable
ground-state wavefunction, can be viewed as the $\alpha \rightarrow 0$
limit of $-{b}/{\alpha \sinh^{2}\left( -x/\alpha +\pi i/3\right) } $
with $\pi a\alpha =6b$. Thus all of the known quantum
mechanical problems with factorisable ground-state wavefunction
are included in (\ref{detdiff}).}
More recently it has been shown \cite{Bra} to characterise the Calogero-Moser
system \cite{vDV}, which is a scaling limit of the Ruijsenaars-Schneider system.
There appear deep connections between functional equations and integrable
systems \cite{Ca2, BCb, inoz, Inoz, BKr, DFS, BB1, BB2, Gur1, Gur2, Gur3, Gur4}.
The analytic solutions to (\ref{detdiff}) were characterised by Buchstaber
and Perelomov \cite{bp} while more recently a somewhat stronger result with
considerably simpler proof was obtained by the authors \cite{BBS}. One has
\begin{thm}[\cite{BBS}]
Let F be a three-times differentiable function satisfying the functional
equation (\ref{detdiff}).
Then, up to the manifest invariance
$ F(z)\rightarrow \alpha F(\delta z)+\beta$,
the solutions of (\ref{detdiff}) are one of $F(z)=\wp(z+d)$,
$F(z)=e\sp{z}$ or $F(z)=z$.
Here $\wp$ is the Weierstrass $\wp$-function and
 $3 d$ is a lattice point of the $\wp$-function.
\end{thm}

Thus the even solutions of (\ref{detdiff}) are precisely those obtained by
Ruijsenaars and Schneider. 
Until this year the general solution to (\ref{functional}) remained unknown
when the authors established 

\begin{thm}[\cite{BBS2}]
The  general even solution of (\ref{functional}) amongst the class of
meromorphic functions whose only singularities on the Real axis are
either a double pole at the origin, or double poles at $n p$ ($p$
real, $n\in \mathbb Z$) is:
\newline a) for all odd $n$ given by the solution of Ruijsenaars and
Schneider while\newline
b) for even $n\ge4$ there
are in addition to the Ruijsenaars-Schneider solutions the following:
\begin{equation*}
\begin{split}
F_1(z)&=\sqrt{(\wp(z)-e_2)(\wp(z)-e_3)}=
\frac{\sigma_2(z)\sigma_3(z)}{\sigma\sp2(z)} \\&=
\frac{\theta_3(v)\theta_4(v)}{\theta_1\sp2(v)}
\frac{\theta_1\sp{\prime2}(0)}{4\omega^2\theta_3(0)\theta_4(0)}
= b\frac{\dn(u)}{\sn\sp2(u)} \\
F_2(z)&=\sqrt{(\wp(z)-e_1)(\wp(z)-e_3)}=
\frac{\sigma_1(z)\sigma_3(z)}{\sigma\sp2(z)} \\&=
\frac{\theta_2(v)\theta_4(v)}{\theta_1\sp2(v)}
\frac{\theta_1\sp{\prime2}(0)}{4\omega^2\theta_2(0)\theta_4(0)}
=b\frac{\cn(u)}{\sn\sp2(u)}\\
F_3(z)&=\sqrt{(\wp(z)-e_1)(\wp(z)-e_2)}=
\frac{\sigma_1(z)\sigma_2(z)}{\sigma\sp2(z)} \\ &=
\frac{\theta_2(v)\theta_3(v)}{\theta_1\sp2(v)}
\frac{\theta_1\sp{\prime2}(0)}{4\omega^2\theta_2(0)\theta_3(0)}
=b\frac{\cn(u)\dn(u)}{\sn\sp2(u)}
\end{split}
\end{equation*}
\end{thm}
Here
$$\sigma_\alpha(z)=
\frac{\sigma(z+\omega_\alpha)}{\sigma(\omega_\alpha)}
e\sp{-z\zeta(\omega_\alpha)},
\qquad u=\sqrt{e_1-e_3}\,z ,
\qquad v=\frac{z}{2\omega},
\qquad b={e_1-e_3}
$$
with $\omega_1=\omega$, $\omega_2=-\omega-\omega'$ and $\omega_3=\omega'$,
and we have given representations in terms of the Weierstrass elliptic
functions, theta functions and the Jacobi elliptic functions \cite{WW}.
For appropriate ranges of $z$ the solutions are real.
Their degenerations yield all the even solutions with only a double pole at
$x=0$ on the real axis.  These degenerations may in fact coincide with
the degenerations of the Ruijsenaars-Schneider solution.
One can straightforwardly verify these new solutions do in fact satisfy
(\ref{functional}) for even $n$ \cite{BB3} but new techniques had to be 
developed to show these exhaust the solutions.

The models discovered by Ruijsenaars and Schneider not only exhibited
an action of the Poincar\'e algebra but were completely integrable as well.
In particular Ruijsenaars and Schneider demonstrated the
Poisson commutativity for their solutions of the light-cone quantities
\begin{equation}
S_{\pm k} =\sum_{\substack{ I\subseteq \{1,2,\ldots, n\}\\ \\ |I|=k} }
\exp\left(\pm {\sum_{i\in I} p_i}\right) \,
\prod_{\substack{ i\in I\\ \\ j\not\in I}} f(x_i-x_j).
\label{lccons}
\end{equation}
Then $H=(S_1+S_{-1})/2$ and $P=(S_1-S_{-1})/2$. 
(Note the even/oddness of the functions $f(x)$ means that there
really are only $n$ functionally independent quantities.)
It is an open problem whether the new solutions of theorem 3 yield
integrable systems.
We know that these new solutions do not always yield Poisson commuting 
quantities using the ansatz of Ruijsenaars and Schneider, but as yet we 
cannot rule out other Poisson commuting conserved quantities \cite{BB3}.

Ruijsenaars \cite{R} also investigated the quantum version of the classical
models he and Schneider introduced. From the outset he sought operator
analogues of the light-cone quantities (\ref{lccons}).
He showed that (for $k=1,\ldots n$)
\begin{equation*}
{\hat S}_{k} =\sum_{\substack{ I\subseteq \{1,2,\ldots, n\}\\ |I|=k} }
\prod_{\substack{ i\in I \\ {j\not\in I}}} h(x_j-x_i)\sp{\frac{1}{2}}\,
\exp\left(-\sqrt{-1}\,\beta {\sum_{i\in I} \partial_i}\right) \,
\prod_{\substack{ i\in I \\ j\not\in I}} h(x_i-x_j)\sp{\frac{1}{2}}
\label{qlccons}
\end{equation*}
pairwise commute if and only if (\ref{qgencons}) held for all $k$ and $n\ge1$.
Further he was able to show  (\ref{eqsol})
led to a solution of (\ref{qgencons}), the solution being related to the
earlier Ruijsenaars-Schneider solution via
$$\frac{\sigma(x+\nu)\sigma(x-\nu)}{\sigma\sp2(x) \sigma\sp2(\nu)}=
\wp(\nu)-\wp(x).
$$
Ruijsenaars \cite{R} suggested that this solution was ``most likely unique"
but was unable to prove this.
A consequence of our classical analysis are the possible functions
$F(x)=h(x)h(-x)$.
A natural question to ask is whether there is a solution to (\ref{qgencons})
corresponding to our new solutions. If not, then the
Ruijsenaars solution is indeed unique.
Our theorem proves the uniqueness of the Ruijsenaars solutions.

We shall now turn to the proof of theorem 1 using the transform method
developed in \cite{BBS2}.

\section{Proof of Theorem 1}

In this section we solve (\ref{qgencons}) by constructing the Fourier 
transform of this equation for $k=1$.
In addition to the function
$F\left(  z\right)  =h\left(  z\right)  h\left(  -z\right)  $
let us introduce
\begin{equation}
g\left( z,\beta\right)  =h\left(  -z\right)  h\left(  z-i\beta\right) .
\label{defg}
\end{equation}
The Fourier transform of (\ref{qgencons}) will be in terms of the Fourier
transform $\widehat{g}\left(  k,\beta\right)$ of $g\left(  z,\beta\right)$.
We will show how, after a judicious choice of parametrization, the
Fourier transform of (\ref{qgencons}) when $k=1$ leads to precisely the
same equation encountered when studying the Fourier transform of
(\ref{functional}). Theorem 3 then gives the general solutions for 
$g\left( z,\beta\right)$. 
We find that we can write
\begin{equation*}
g\left(  z,\beta\right)  \equiv h\left(  -i\beta\right)  
\Big(  F_{1}\left( z-i\beta\right)  -F_{1}\left(  z\right)  
-F_{1}\left(  \nu-i\beta\right)  +F_{1}\left(  \nu\right)\Big)
\end{equation*}
where $F_{1}\left(  z\right)  =\int F\left(  z\right)  dz$, $F(\nu)=0$
and $F(x)$ is any solution given in theorem 3.
Now given $g(z,\beta)$ we wish to factorise this in the form (\ref{defg}).
The new solutions of theorem 3 do not
have the factorisation property (\ref{defg}), whereby we establish that
for $k=1$ the only solutions to (\ref{qgencons}) are given by (\ref{eqsol}).
It is known however that these solutions satisfy (\ref{qgencons}) for all
$k$, and so the theorem will be proved.
Our strategy will be to take the  Fourier transform for functions 
of increasing complexity,
first considering those functions with only a pole at the origin and vanishing
at infinity; next we consider similar functions decaying to a constant at
infinity; finally we consider those functions with a periodic array of
poles along the real axis including the origin.

Before we derive the equation for $\widehat{g}$ we look at the properties of 
$g$ and the conditions that these properties demand of $\widehat{g}.$ The 
original solutions of Ruijsenaars and Schneider for $F\left(  z\right)  $ can 
be expressed as
\begin{equation}
F\left(  z\right)  =A\wp\left(  z,g_{2},g_{3}\right)  +B, 
\label{3.1}
\end{equation}
where $A$, $B$, $g_{2}$ and $g_{3}$ are constants. These are 
even functions of $z$
with a double pole at the origin. The new solutions of \cite{BBS2}
have a similar structure but do not possess the addition of the arbitrary
constant $B$. However, in both cases we can use the constant $A$ and the scaling
properties of $\wp\left(  z\right)  $ to restrict our choice of $F\left(
z\right)  ,$ without loss of generality, so that $z^{2}F\left(  z\right)
\rightarrow-1$ as $z\rightarrow0$ and that when the solution has finite real
period we take this to be $2\pi$. Let $\nu$ be a zero of $F(z)$. Thus for the
solutions (\ref{3.1}) we may express the constant $B$ in terms of $\nu$ as
\begin{equation}
F\left(  z\right)  =\wp\left(  \nu\right)  -\wp\left(  z\right)  . 
\label{3.2}
\end{equation}
The condition $F(\nu)=0$ then requires
$h\left(  \nu\right)  h\left(  -\nu\right)  =0$.
We fix $h\left(  z\right) $ by demanding that $h\left(z\right)=0$ at 
$z=-\nu$. This entails
$g\left(  \nu,\beta\right)  =g\left(  -\nu+i\beta,\beta\right)  =0.$ Also
since $F\left(  z\right)  $ has a double pole at $z=0,\;h\left(  z\right)  $
must have a simple pole at $z=0$ and we choose $h(z)$ so that
$zh(z)\rightarrow+1$ as $z\rightarrow0.$ Thus $g\left(  z,\beta\right)  $
must have simple poles at $z=0$ and $z=i\beta$ with $zg\left(  z\right)
\rightarrow-h\left(-  i\beta\right)  $ as $z\rightarrow0$ and $\left(
z-i\beta\right)  g\left(  z\right)  \rightarrow h\left(  -i\beta\right)  $ as
$z\rightarrow i\beta.$

We will now obtain an equation for $\widehat{g}$ by taking the
Fourier transform of (\ref{qgencons}). Set $z_j=x_{j}+iy_{j}$ and
denote by $E(  \mathbf{k},z_{n})$ the $k=1$ equation (\ref{qgencons}),
where
$\mathbf{z}$ is the vector $\left(  z_{1},z_{2},...z_{n-1}\right)$.
We define the $(n-1)$-dimensional Fourier
transform by
\begin{equation}
\widehat{E}\left(  \mathbf{k},z_{n},\beta\right)  =
\int\limits_{\mathbb{R}^{n-1}}E\left(\mathbf{z},z_{n},\beta\right) 
 e^{-i\mathbf{k.z}}d\mathbf{z}. 
\label{3.3}
\end{equation}
However, since $g\left(  z,\beta\right)  $ has a pole at the origin, we
replace $z_{j}$ by $z_{j}+i\epsilon_{j}$ and assume that $\epsilon_1>
\epsilon_2>\ldots>\epsilon_{n}>0$ and that $\epsilon_{1}$ is small. We
then assume in the definition of $\widehat{E}$ in (\ref{3.3}) that we integrate
along the Real axis in the complex $x_{j}+iy_{j}$ plane. 
\subsection{Functions with infinite real period vanishing at infinity}
In the first
instance we consider the class of solutions $g\left(  z,\beta\right)  $ which
have infinite real period and tend to zero at infinity with no other
singularities on the real axis other than the pole at the origin. Hence, when
$z_{j}$ has been replaced by $z_{j}+i\epsilon_{j}$, the integrand will have no
other singularities in the domain of integration provided $\epsilon_{1}<\beta$ 
which we take to be real and positive. The reduction of
$\widehat{E}=0$ as $\epsilon_{1}\rightarrow0$, to an equation involving the
generalised Fourier transform $\widehat{g}(k,\beta)$ follows the lines of
\cite{BBS2}. The definition of $\widehat{g}$ is given by
\begin{equation}
\widehat{g}\left(k,\beta\right)  =\frac{1}{2}\left(  \widehat{g}_{U}\left(
k,\beta\right)  +\widehat{g}_{L}\left(  k,\beta\right)  \right)  , 
\label{3.4}
\end{equation}
where
\begin{equation}
\widehat{g}_{U}\left(  k,\beta\right)  =\int\limits_{-\infty}^{\infty}
\!\!\!\!\!\!\!\cap g\left(  z,\beta\right)  e^{-ikz}dz, 
\qquad
\widehat{g}_{L}\left(  k,\beta\right)  =\int\limits_{-\infty}^{\infty}
\!\!\!\!\!\!\!\cup g\left(  z,\beta\right)  e^{-ikz}dz, 
\label{3.6}
\end{equation}
are defined respectively to go over and under the pole at $z=0$.
Then we have
\begin{equation*}
\widehat{g}_{L}-\widehat{g}_{U} =2\pi i\times\text{ Residue }g\left(
z,\beta\right)  |_{z=0} =-2\pi ih\left(  -i\beta\right)  
\label{3.7}
\end{equation*}
and 
$$ \widehat{g}_{U}=\widehat{g}+i\pi h(-i\beta),\qquad 
\widehat{g}_{L}=\widehat{g}-i\pi h(-i\beta).$$

However the above definition causes a problem as $\beta\rightarrow0$ in that
\begin{equation}
\underset{\beta\rightarrow0}{\lim}\ \widehat{g}\left(  k,\beta\right)
\neq\widehat{g}\left(  k,0\right)  . \label{3.8}
\end{equation}
This is because when $\beta>0$ both paths of integration in (\ref{3.6}) 
lie below the pole at $z=i\beta,$ while if we put $\beta=0$ 
and then use the definitions (\ref{3.4}) and (\ref{3.6}) we find that the 
upper path of integration will go above the pole which becomes a double pole 
at $z=0$ as $\beta\rightarrow0$. To overcome this difficulty we define a 
modified $\widehat{g}_{U}$, $\widehat{g}_{UM}$, and a modified generalised 
Fourier transform $\widehat{g}_{M},$ by indenting the upper
contour in (\ref{3.6}) so that it goes over the pole at $z=i\beta$. Then
\begin{equation*}
\widehat{g}_{UM} =\widehat{g}_{U}-2\pi i\;\text{residue of }g(z,\beta)
e\sp{-ikz}\big|_{z=i\beta}
=\widehat{g}_{U}-2\pi ih\left(  -i\beta\right)  e^{\beta k}.\label{3.9}
\end{equation*}
Hence
\begin{equation}
\widehat{g}_{M}=\widehat{g}-\pi ih\left(  -i\beta\right)  e^{\beta k}.
\label{3.10}
\end{equation}
With these definitions it is easy to show that when $g=-1/(z\left(
z-i\beta\right)),$ $\widehat{g}_{M}=\pi(e^{\beta k}-1)$
sign$(k)/\beta$ with $\lim\limits_{\beta\rightarrow0}\widehat{g}_{M}=\pi\left|
k\right|$, which is the generalised Fourier transform of $-1/z^{2}$.

When we take the Fourier transform of (\ref{qgencons}) for $k=1$
each of the terms $\prod_{j\ne i} h(x_j-x_i) h(x_i-x_j-i\beta)=
\prod_{j\ne i}g(x_i-x_j)$ in the sum reduces to a product of one dimensional 
Fourier transforms. Using the above definitions, one of four
possibilities arise depending on the position of the poles:
\begin{align*}
\int_{-\infty}\sp{\infty}g(u+i\epsilon)e\sp{-iku}du&=
\widehat{g}_{M}(k)+i\pi h\left(-i\beta\right)-i\pi h(-i\beta)e^{\beta k},
\\
\int_{-\infty}\sp{\infty}g(u-i\epsilon)e\sp{-iku}du&=
\widehat{g}_{M}(k)-i\pi h\left(-i\beta\right)+i\pi h(-i\beta)e^{\beta k},
\\
\int_{-\infty}\sp{\infty}g(-u+i\epsilon)e\sp{-iku}du&=
\widehat{g}_{M}(k)+i\pi h\left(-i\beta\right)-i\pi h(-i\beta)e^{-\beta k},
\\
\int_{-\infty}\sp{\infty}g(-u-i\epsilon)e\sp{-iku}du&=
\widehat{g}_{M}(k)-i\pi h\left(-i\beta\right)+i\pi h(-i\beta)e^{-\beta k}.
\end{align*}
Thus for example (relevant to the $n=3$ case)
\begin{align*}
\widehat{g(z_1-z_2) g(z_1-z_3)}&=
\int_{{\mathbb R}\sp2}g(z_1-z_2+i(\epsilon_1-\epsilon_2))
g(z_1-z_3+i(\epsilon_1-\epsilon_3))e\sp{-ik_1 z_1 -ik_2 z_2}dz_1 dz_2\\
&=e\sp{-i(k_1+k_2)z_3}
\int_{-\infty}\sp{\infty}dv g(v+i\epsilon')e\sp{-i(k_1+k_2)v}
\int_{-\infty}\sp{\infty}du g(-u+i\epsilon'')e\sp{-ik_2u}\\
&=e\sp{-i(k_1+k_2)z_3}
\big[ \widehat{g}_{M}(k_1+k_2)+i\pi h\left(-i\beta\right)-
i\pi h(-i\beta)e^{\beta (k_1+k_2)} \big]\\
&\qquad \times
\big[ \widehat{g}_{M}(-k_2)+i\pi h\left(-i\beta\right)-
i\pi h(-i\beta)e^{-\beta k_2} \big].
\end{align*}
Here we have set $u=z_2-z_1$, $v=z_1-z_3$, $\epsilon'=\epsilon_1-\epsilon_3$
and $\epsilon''=\epsilon_1-\epsilon_2$.
The common factor of $h(-i\beta)$ in these expressions suggests the rewriting 
$\widehat{g}_{M}\left(  k,\beta\right)=-iI\left(k,\beta\right)\beta h\left(
 -i\beta\right)$. Then in the limit $\beta\rightarrow0$, 
$\widehat{g}_{M}\left(  k,0\right)  =I\left( k,0\right)$. 
If we use this we find that the $n=3$ equation (\ref{3.3}) can be written as
$\sum\limits_{1}^{3}J_{j}=0$, where for example
\begin{align}
J_{1}  &  =-\left(  I\left(  k_{1}+k_{2},\beta\right) \frac{\beta}{\pi}
-1+e^{\beta \left( k_{1}+k_{2}\right)  }\right) 
 \left(  I\left(  -k_{2},\beta\right)
\frac{\beta}{\pi}-1+e^{-\beta k_{2}}\right) \nonumber\\
&\quad  +\left(  I\left(  -k_{1}-k_{2},\beta\right) \frac{\beta}{\pi}
+1-e^{-\beta\left( k_{1}+k_{2}\right)}\right)\left(I\left(k_{2},\beta\right)
\frac{\beta}{\pi} +1-e^{\beta k_{2}}\right)  . 
\label{3.11}
\end{align}
This corresponds to $\widehat{g(z_1-z_2) g(z_1-z_3)}-
\widehat{g(z_2-z_1) g(z_3-z_1)}$ in the sum, 
with similar definitions for $J_{2}$ and $J_{3}$:
\begin{align*}
J_{2}  &  =\left(  I\left( - k_{1}-k_{2},\beta\right) \frac{\beta}{\pi}
+1-e^{-\beta \left( k_{1}+k_{2}\right)  }\right)
 \left(  I\left(  k_{1},\beta\right)
\frac{\beta}{\pi}-1+e^{\beta k_{1}}\right) \nonumber\\
&\qquad  -\left(  I\left(  k_{1}+k_{2},\beta\right) \frac{\beta}{\pi}
-1+e^{\beta\left( k_{1}+k_{2}\right)}\right)\left(I\left(-k_{1},\beta\right)
\frac{\beta}{\pi} +1-e^{-\beta k_{1}}\right)  ,\\
J_{3}  &  =\left(  I\left( k_{1},\beta\right) \frac{\beta}{\pi}
-1+e^{\beta  k_{1} }\right)
 \left(  I\left(  k_{2},\beta\right)
\frac{\beta}{\pi}-1+e^{\beta k_{2}}\right) \nonumber\\
&\qquad  -\left(  I\left(-  k_{1},\beta\right) \frac{\beta}{\pi}
+1-e^{-\beta k_{1}}\right)\left(I\left(-k_{2},\beta\right)
\frac{\beta}{\pi} +1-e^{-\beta k_{2}}\right). 
\end{align*}

At this point we introduce what, with hindsight, will prove a judicious
change of variable: set
\begin{equation}
I\left(  k,\beta\right)  =\widehat{G}\left(  k,\beta\right)  \left(  e^{\beta
k}-1\right)  /\left(  k\beta\right). \label{3.12}
\end{equation}
This change of variable is suggested by a careful examination of the series for 
$I\left(k,\beta\right)=\sum_{0}^{\infty}I_{j}\left(k\right)\beta^{j}$
that results from $\sum\limits_{1}^{3}J_{j}=0$ with the expressions above.
Whatever, when $\beta\rightarrow0$, 
$I\rightarrow \widehat{G}\left(k,0\right)$ and we know (as $F(x)$ is even)
that $\widehat{G}\left( k,0\right)$ is even. An examination of the equation for 
$\widehat{G}\left(k,\beta\right)$ (say by a series expansion) shows further
that $\widehat{G}\left(k,\beta\right)$ itself is even. This fact leads to
simplifications. When $\widehat{G}\left(  k,\beta\right)$ is even
$J_{1}$ simplifies to
\begin{equation*}
J_{1}  =-\frac{\left(  e^{\beta k_{1}}-1\right)  \left(  e^{\beta k}-1\right)
\left(  1-e^{-\beta\left(  k_{1}+k_{2}\right)  }\right) }{(k_{1}+k_{2})k_{2}} 
\, \Big(\widehat{G}\left( k_{2},\beta\right) -\pi k_{2}\Big)
 \left(  \widehat{G}\left(  k_{1}+k_{2},\beta \right)  +\pi(k_{1}+k_{2})
\right), 
\label{3.13}
\end{equation*}
and similarly for $J_2$, $J_3$.
We then find the  exponential factors involving $\beta$ are common to all 
$J_{i}$ and so (\ref{3.3}) can finally be expressed, in the case $n=3$ as
\begin{equation}
\left(  k_{2}\widehat{G}\left(  k_{1},\beta \right)  +k_{1}\widehat{G}\left(
k_{2},\beta \right)  \right)  \widehat{G}(k_{1}+k_{2},\beta ) 
-\left(  k_{1}+k_{2}\right)
\widehat{G}\left(  k_{1},\beta \right)  \widehat{G}(k_{2},\beta )
=\pi^{2}k_{1}k_{2}\left(  k_{1}+k_{2}\right)  . \label{3.14}
\end{equation}

Now this is precisely the equation for the Fourier transform of 
$F\left(  z\right)$ obtained in \cite{BBS2}[eqn. 5.5]
when studying (\ref{functional})
for $n=3$. The only solutions of the required from are 
$\widehat{G}\left(k,\beta \right)\equiv\widehat{G}\left( k\right) 
=\widehat{F}\left(  k\right)  =\pi k\coth\left(  \pi k/a\right)$
and its limit as $a\rightarrow0$ namely $\pi\left|  k\right|$.
Hence
\begin{equation}
\widehat{g}_{M}\left(  k,\beta\right)  =-i\beta h\left(  -i\beta\right)
\frac{\left(  e^{\beta k}-1\right)  }{\beta k}\widehat{F}\left(  k\right)  ,
\label{3.15}
\end{equation}
or
\begin{equation}
g\left(  z,\beta\right)  =h\left(  -i\beta\right)  \left(  \int F\left(
z-i\beta\right)  dz-\int F\left(  z\right)  dz\right)  . \label{3.16}
\end{equation}
It is then easy to check that for
\begin{equation}
\widehat{F}\left(  k\right)  =\pi k\coth\left(  \pi k/a\right),
\qquad F\left( z\right)=-\frac{1}{4}a^{2}/\sinh^{2}\left(\frac{1}{2}az\right),
\label{3.17}
\end{equation}
we have
\begin{align}
g\left(  z,\beta\right) &=\frac{a}{2}h\left(  -i\beta\right)  \left(
\coth\frac{1}{2}a\left(  z-i\beta\right)  -\coth\left(  \frac{1}{2}az\right)
\right) \nonumber\\
& =\frac{a}{2}h\left(  -i\beta\right)\frac{\sinh\left( \frac{1}{2}
 ia\beta\right) }
{\sinh\left(  \frac{1}{2}az\right)  \sinh\frac{1}{2}a\left(  z-i\beta\right)
}\equiv h\left(  -z\right)  h\left(  z-i\beta\right),
\label{3.18}
\end{align}
with $h\left(z\right)  =\frac{1}{2}a/\sinh\left(  \frac{1}{2}az\right)  $
giving the appropriate factorisation
$ F\left(  z\right)  =h\left(  -z\right)  h\left(  z\right)$.

Similarly (or let $a\rightarrow0$) for $\widehat{F}\left(  k\right)  =\pi|k|$
and $F\left(  z\right)  =-1/z^{2}$ we have
\begin{equation}
g\left(  z,\beta\right) =h\left(  -i\beta\right)  \left(  \frac
{1}{z-i\beta}-\frac{1}{z}\right) 
=h\left(  -i\beta\right)  \frac{-i\beta}{\left(  -z)(z-i\beta\right)}
\equiv h\left(  -z\right)  h\left(  z-i\beta\right),
\label{3.20}
\end{equation}
with $h\left(  z\right)  =1/z$ and 
$F\left(  z\right)  =h\left( -z\right)  h\left(  z\right)$.
Both these factorisations are only defined up to the shift by the exponential
$h(z)\rightarrow h(z)e\sp{\alpha z}$ given in (\ref{eqsol}).

For the cases $n\geq4$ we also find that the same set of transformations of
$\widehat{g}_M$ reduce equation (\ref{3.3}) to the original equation for the 
transform $\widehat{F}$ encountered in the study of (\ref{functional}).
For odd $n$ we only have the solutions (\ref{3.16}) 
and (\ref{3.20}) while for even $n$ we have in addition to these solutions
the solution
\begin{equation}
\widehat{F}\left(  k\right)  =\pi k\tanh\left(  \frac{\pi k}{2 a}\right),
\qquad F\left(  z\right)  =-\frac{a^{2}\cosh az}{\sinh^{2}az}. \label{3.22}
\end{equation}
Now however we have
\begin{equation}
g\left(  z,\beta\right)  =a h\left(  -i\beta\right) \left(  \frac{1}{\sinh
a\left(  z-i\beta\right)  }-\frac{1}{\sinh az}\right), 
\label{3.23}
\end{equation}
which cannot be written in the form 
$g\left(  z,\beta\right) =h\left(  -z\right)  h\left(  z-i\beta\right)$. Thus
these do not yield solutions to (\ref{qgencons}). 

\subsection{Functions with infinite real period constant at infinity}
The solutions (\ref{3.1}) of Ruijsenaars and Schneider contain the addition 
of an arbitrary constant that, in the hyperbolic limit, corresponds to
the function not vanishing at infinity.
The new solutions of \cite{BBS2}, of which (\ref{3.22}) is an example, do not
have this degree of freedom:
the hyperbolic degenerations of these solutions all tend to  zero at infinity.
To deal with functions which do not tend to zero at infinity we must deal with
distributional Fourier transforms. 
The addition of a constant to $F\left(z\right)$ requires an addition of a 
similar constant to $g\left(z,\beta\right)$, since for example if  
$F \to a^2$ as $z\to\infty$ then $g \to a\sp2$ as $z\to\infty$. 
For $n=3$ we can easily verify that 
$g\left(z,\beta\right) \rightarrow g\left( z,\beta\right) +$ constant leaves
(\ref{qgencons}) invariant.  
However, this is not automatically the case when $n >3$.  
We find that for $n >3$ the transformation $g\left(
z,\beta\right) \rightarrow g\left( z,\beta\right) +A$, requires extra
conditions on $g$ to leave the equation invariant.  When $n=4$, there
is only one extra condition which is automatically satisfied for the
solution $g$ corresponding to the solution $F$ given by Ruijsenaars and
Schneider, but not for the solution $F$  corresponding to the new solutions
of \cite{BBS2}.  We believe that for $n >4$, the
Ruijsenaars and Schneider solution automatically satisfy all the extra
conditions but that the new solutions given in theorem 3 do not. 
The addition of an arbitrary constant $A$, to $g$, requires the addition
of $A\delta\left( k\right) $ to $\widehat{g}(k,\beta)$. For the solution
$\widehat{g}$ which are otherwise well behaved functions of $k$ we can
easily obtain the solution for $g\left( z,\beta\right) $ and the
corresponding solution for $h\left(  z\right)$, when
they exist. For example, instead of (\ref{3.17}) we add an arbitrary constant
to $F\left(  z\right)$ and, demanding that $F\left(\nu\right)=0$, we have
\begin{equation}
F\left(  z\right)  =\frac{1}{4}a^{2}\left\{  \frac{1}{\sinh^{2}\left(
\frac{1}{2}a\nu\right)  }-\frac{1}{\sinh^{2}\left(  \frac{1}{2}az\right)
}\right\}  , \label{3.23b}
\end{equation}
so that we have
\begin{equation}
g\left(  z,\beta\right)  =\frac{1}{2}ah\left(  -i\beta\right)  \frac
{\sinh\left(  -\frac{1}{2}ai\beta\right)  }{\sinh\left(  -\frac{1}
{2}az\right)  \sinh\frac{1}{2}a\left(  z-i\beta\right)  }+A. \label{3.24}
\end{equation}

Since by definition $g$($z,\beta)=h\left(  -z\right)  h\left(  z-i\beta
\right)  $ and $h\left(  z\right)  $ has a zero at $z=-\nu$, we have  $g\left(
\nu,\beta\right)  =0,$ as indicated earlier.  Hence
\begin{equation}
A\equiv A\left(  \beta,\nu\right)  =-\frac{1}{2}ah\left(  -i\beta\right)
\frac{\sinh\left(  -\frac{1}{2}ai\beta\right)  }{\sinh\left(  -\frac{1}{2}
a\nu\right)  \sinh\frac{1}{2}a\left(  \nu-i\beta\right)  }. \label{3.25}
\end{equation}
Then we may express $g$ as
\begin{align}
g\left(  z,\beta\right) &=\frac{1}{2}ah\left(  -i\beta\right) 
\frac{\sinh \frac{1}{2}a\left(  -z+\nu\right)  }{\sinh(\frac{1}{2}a\nu)\sinh\left(  -\frac{1}{2}az\right)  }
\frac{\sinh \frac{1}{2}a\left(z+\nu-i\beta\right)  }{\sinh \frac{1}{2}a\left(  z-i\beta\right)  }
\frac{\sinh\left( -\frac{1}{2}ai\beta\right)  }{\sinh\frac{1}{2}a\left(  \nu-i\beta\right)}
\nonumber\\
&\equiv h\left(  -z\right)  h\left(  z-i\beta\right), 
\label{3.26}
\end{align}
where
\begin{equation}
h\left(  z\right)  =\frac{1}{2}a\frac{\sinh\frac{1}{2}a\left(  z+\nu\right)
}{\sinh\left(  \frac{1}{2}az\right)  \sinh(\frac{1}{2}a\nu)}, \label{3.27}
\end{equation}
with
\begin{equation}
h\left(  -z\right)  h\left(  z\right) =\frac{a^{2}}{4}\frac{\sinh
\frac{1}{2}a\left(  z-\nu\right)  \sinh \frac{1}{2}a\left(  z+\nu\right)  }{\sinh^{2}\left(
\frac{1}{2}az\right)  \sinh^{2}\left(  \frac{1}{2}a\nu\right)  }
=\frac{a^{2}}{4}\left\{  \frac{1}{\sinh^{2}\left(  \frac{1}{2}a\nu\right)
}-\frac{1}{\sinh^{2}\left(  \frac{1}{2}az\right)  }\right\}, 
\label{3.28}
\end{equation}
as required.
Also observe that as $\nu\rightarrow \infty$ we have $h(z)$ tending to the
solutions of the previous section times the exponential factor 
$e\sp{az/2}$, and our factorisation is only unambiguous up to such terms.

A similar but easier calculation for $F\left(  z\right)  =1/\nu^{2}-1/z^{2}$
gives
\begin{equation}
g\left(  z,\beta\right)  =h\left(  -i\beta\right)  \frac{\nu\left(
-i\beta\right)  }{\nu-i\beta}\frac{\left(  -z+\nu\right)  }{\left(  -z\right)
\nu}\frac{z+\nu-i\beta}{\left(  z-i\beta\right)  \nu}, \label{3.29}
\end{equation}
so that $h\left(  z\right)  =\left(  z+\nu\right)  /\left(  z\nu\right)  $
with $h\left(  z\right)  h\left(  -z\right)  =1/\nu^{2}-1/z^{2}.$

\subsection{Periodic functions}
For $2\pi$ periodic functions which have a periodic array of double poles at
$z=2\pi p$ we have shown in \cite{BBS2} that the appropriate form of the 
transform $\hat{g}\left(k,\beta\right)$ is $\sum\limits_{p=-\infty}^{\infty}
a_p\delta\left(k-p\right)$ corresponding to a Fourier series $\frac{1}
{2\pi}\sum\limits_{p=-\infty}^{\infty}a_pe^{ipz}$ for $g\left(z,\beta\right)$.
The equation satisfied by $\hat{g}$ is the same as that for the
non-periodic case, but it is solved only at integer values of the $\left\{
k_{j}\right\}$. Thus, if $\hat{g}\left(  k,\beta\right)$ is the solution
of the continuous case we have the solution $a_{p}\left(\beta\right)
=\hat{g}\left(p,\beta\right)$.  The corresponding solution, $g_{2\pi
}\left(  z,\beta\right)$, is the periodic extension of the continuous case
expressed in the form
\begin{equation}
g_{2\pi}\left(  z,\beta\right)  =\sum\limits_{p=-\infty}^{\infty}g\left(  z-2\pi
p\right) \label{3.30}
\end{equation}
where $g(z)$ is the nonperiodic solution determined  by (\ref{3.16})
with the particular forms (\ref{3.18}), (\ref{3.20}), (\ref{3.23}), 
(\ref{3.26}) and (\ref{3.29}).

For the function $F(z)=1/\sinh^{2}z$ this corresponds to the $\wp$ function. 
Since without loss of generality we have taken $F\left(  z\right)  $ to satisfy
$z^{2}F\left(  z\right)  =-1$ as $z\rightarrow0,$ we can use the scaling
property $\wp\left(  z,g_{2},g_{3}\right)  =a^{2}\wp\left(  az,g_{2}
/a^{4},g_{3}/a^{6}\right)  $ to take $F\left(  z\right)  =-\wp\left(
z\right)  .$ The addition of a constant to $F$ gives rise to the addition of
a constant to $g,$ thus
\begin{equation}
F\rightarrow F+B\Rightarrow\int F\rightarrow\int F+Bz+B_{1}\Rightarrow
g\rightarrow g-iB\beta h\left(  -i\beta\right)  .\label{3.31}
\end{equation}
Hence a periodic solution corresponding to (\ref{3.18}) is
\begin{equation}
g\left(  z,\beta\right)  =-h\left(  -i\beta\right) \Big(  \zeta\left(
z\right)  -\zeta\left(  z-i\beta\right)  +C\left(  \beta\right)  \Big)
\label{3.32}
\end{equation}
where again $C\left(\beta\right)$ is an arbitrary function of $\beta$, which
also depends on the parameter $\nu$, and is determined by the condition
$g\left(  \nu,\beta\right)=0$. Thus
\begin{align}
g\left(  z,\beta\right) &  
=-h\left(  -i\beta\right)  \left\{  \zeta\left( z\right)  -
  \zeta\left(  z-i\beta\right)  -\zeta\left(  \nu\right)  +
  \zeta\left(\nu-i\beta\right)  \right\} \nonumber\\
&  =h\left(  -i\beta\right)  \frac{\sigma\left(  \nu\right)  \sigma\left(
-i\beta\right)  }{\sigma\left(\nu-i\beta\right)  }\frac{\sigma\left(
-z+\nu\right)  }{\sigma\left(  -z\right)  \sigma(\nu)}\frac{\sigma\left(
z+\nu-i\beta\right)  }{\sigma\left(  z-i\beta\right)  \sigma\left(
\nu\right)}
\equiv h\left(  -z\right)  h\left(  z-i\beta\right)  \label{3.33}
\end{align}
with 
$h\left(z\right)=\sigma\left(z\right)=\sigma\left(z+\nu\right)
/\left(\sigma\left(z\right)\sigma\left(\nu\right)\right)$.

Again the solution for $g$ for the new solutions of Byatt-Smith and
Braden \cite{BBS2} can be written in a form similar to (\ref{3.32}) but 
again cannot be factorised into the product
$h\left( -z\right) h\left(z-i\beta\right)$.
For example consider
$F(z)=-\mbox{cn}(z)\mbox{dn}(z)/\mbox{sn}(z)^2$, whence 
$ \int F(z)dz=1/\mbox{sn}(z)$.  Thus from (\ref{3.16})
\begin{equation}
g\left(  z,\beta\right)  =h\left( -i\beta\right)(1/\mbox{sn}(z-i\beta) 
-1/\mbox{sn}(z) )
\end{equation}
and a series approach shows that this cannot be written in factorised
form.

\subsection{Another Functional Equation}
At this stage we have proven the theorem. We have shown that the
Fourier transform of (\ref{qgencons}) with $k=1$ leads to
studying
\begin{equation}
g\left(  z,\beta\right)  \equiv h\left(  -i\beta\right)  \left(  F_{1}\left(
z-i\beta\right)  -F_{1}\left(  z\right)  +C\left(  \beta\right)  \right)
=h\left(  -z\right)h\left(z-i\beta\right).
\label{3.34}
\end{equation}
Here $F_{1}\left(  z\right)  =\int F\left(  z\right)  dz$ and $C\left(
\beta\right)  =-F_{1}\left(  \nu-i\beta\right)  +F_{1}\left(  \nu\right)$,
since we stipulated $g(\nu,\beta)=0$. We found that $F(z)$ had to
be a solution given by theorem 3, and that amongst these the only 
such solutions allowing the desired factorisation  were given by (\ref{eqsol}).
We shall conclude by showing that an analysis of (\ref{3.34}) directly
yields this result.

We view (\ref{3.34}) as a functional equation for $C$, $h$ and $F_{1}$,
with $g$ being consequently determined. We solve this subject to appropriate 
conditions that we inherit from the original problem:
\begin{equation}
zh\left(  z\right)  \rightarrow a,\qquad zF_{1}\left(  z\right)
\rightarrow a,\quad\text{and}\quad g(\nu,\beta)=0.
\label{3.35}
\end{equation}
The last condition means that
$C\left(\beta\right)=-F_{1}\left(\nu-i\beta\right) +F_{1}\left(\nu\right)$,
and so $C(0)=0$.  We observe that (\ref{3.34}) is invariant under
$$
F_1(z)\rightarrow F_1(z)+Bz+B_1,\qquad C(\beta)\rightarrow C(\beta)+
i\beta B.$$
We may fix this freedom by further requiring that
\begin{equation}
C\left(  0\right)  =C^{\prime}\left(  0\right)  =0.\label{3.36}
\end{equation}

To solve (\ref{3.34}) we expand the equation as a power series in $\beta$. The
first  non-trivial terms give sufficient equations to eliminate $h\left(
z\right)  $ and $h\left(  -z\right)  $ and derive a third order equation for
$F_{1}\left(  z\right)  $ in terms of the coefficients $\left\{  C_{0}
,C_{1},b_{0},b_{1},b_{2}\right\}  $ in the expansions
\begin{equation}
C\left(  \beta\right)  =\sum_{j=0}^{\infty}C_{j}\beta^{j+2
}\;\text{and }h\left(  -i\beta\right)  =\sum_{j=0}^{\infty}b_{j}\left(
-i\beta\right)  ^{j-1}. \label{3.37}
\end{equation}
The resulting third order equation for $F_{1}$,
\begin{equation}
 (2F_1^{\prime \prime \prime }(x)
F_1^{\prime }(x) - 3{F_1^{\prime \prime }(x)}^{2})b_{0}^{2}
+ 12C_{0}^{2}b_{0}^{2}
+12{F_1^{\prime }(x)}^{2}(b_{1}^{2} -2b_{2}b_{0})
+24iC_{1}b_{0}^{2}F_1^{\prime }(x)=0,
\end{equation}
can be integrated to give
\begin{equation}
F_{1}\left(  z\right)  =b_{0}\,\zeta\left(  z,g_{2},g_{3}\right)  +z\,b_{0}
\,\wp\left(  \nu,g_{2},g_{3}\right)  , \label{3.38}
\end{equation}
with $C$ determined as
\begin{equation}
C\left(  \beta\right)  =b_{0}\,\zeta\left(  \nu\right)  -b_{0}\,\zeta\left(\nu
-i\beta\right)  +i\beta\, b_{0}\,\wp\left(  \nu\right)  . \label{3.39}
\end{equation}
(The remaining constants in this are defined in terms of
$C_{0},C_{1},b_{1}$ and $b_{2}$ below.)

The function $h\left(  z\right)  $ is then determined by the equation
\begin{equation}
\frac{h^{\prime}}{h}=\frac{1}{2}\frac{F_{1}^{\prime\prime}}{F_{1}^{\prime}}
+\frac{1}{2}b_{0}\wp^{\prime}\left(  \nu\right)  +\frac{b_{1}}{b_{0}}, \label{3.40}
\end{equation}
with solution
\begin{equation}
h(z)=b_0\, \frac{\sigma\left(  z+\nu\right) }{\sigma\left(  z\right)
\sigma\left(  \nu\right)  } e^{\alpha z}. \label{3.41}
\end{equation}

The four constants $\left\{  C_{0},C_{1},b_{1},b_{2}\right\}  $ define the
four constants appearing in (\ref{3.38}-\ref{3.39}) namely 
$\wp\left(  \nu\right)$, $\xi\left(\nu\right)$, $g_{2}$ and 
$\wp'\left(\nu\right)$, with $g_{3}$ given by 
$\left\{\wp\left(\nu\right), \wp'\left(\nu\right)\;\text{and }g_{2}\right\}$.
Of course, to satisfy (\ref{3.35}) we also require $b_{0}=a$.
The relations between the two sets of constants are given by
\begin{equation*}
\begin{array}{ll}
C_{0}=-\frac{1}{2}b_{0}\wp'\left(  \nu\right),&
C_{1}=ib_{0}\left(  \wp\left(  \nu\right)  -\frac{1}{6}g_{2}\right),\\
b_{1}=b_{0}\left(  \zeta\left(  \nu\right)  +\alpha\right),\qquad & 
b_{2}=\frac{1}{2}b_{0}\left(  \left(  \zeta\left(  \nu\right)  +\alpha\right)
^{2}-\wp\left(  \nu\right)  \right) . 
\end{array}
\end{equation*}
The constant $\alpha$ is arbitrary and does not affect the solution (\ref{3.38})
since it is clear that the quotient
$h\left(-z\right)  h\left(z-i\beta\right)/h\left(-i\beta\right)$ is 
independent of $\alpha$. This is the ambiguity in the factorisation
noted earlier.

We conclude that only the solutions of Ruijsenaars and Schneider,
given by (\ref{eqsol}), yield solutions of (\ref{3.34}) and consequently
solutions of (\ref{qgencons}). Thus Theorem 1 is again proved.

\section{Acknowledgements}
We wish to thank A. M. Davie and S.N.M. Ruijsenaars for helpful discussion.
One of the authors (H.W.B.) wishes to thank the Newton Institute for
support during the completion of this work.

\end{document}